\documentclass[
    ,final            
  ]
  {aipproc}

\layoutstyle{8x11double}
\newcommand{\Msun}{\ifmmode {M_{\odot}}\else${M_{\odot}}$\fi}

\begin{document}

\title{SAX J1808.4-3657 in Quiescence: A Keystone for Neutron Star Science}


\author{Heinke, C.~O.}{
  address={University of Virginia, Dept. of Astronomy, P.O. Box 400325, Charlottesville, VA 22904-4325}
,altaddress={Northwestern University, Dept. of Physics \&
  Astronomy, 2145 Sheridan Rd., Evanston, IL 60208} 
}

\author{Deloye, C.~J.}{
  address={Northwestern University, Dept. of Physics \&
  Astronomy, 2145 Sheridan Rd., Evanston, IL 60208}
}

\author{Jonker, P.~G.}{
  address={SRON, Netherlands Institute for Space Research, Sorbonnelaan 2, 3584~CA, Utrecht, the Netherlands}
  ,altaddress={Harvard--Smithsonian  Center for Astrophysics, 60 Garden Street, Cambridge, MA~02138, MA, U.S.A.} 
  ,altaddress={Astronomical Institute, Utrecht University, PO Box 80000, 3508 TA, Utrecht, the Netherlands} 
}

\author{Taam, R.~E.}{
  address={Northwestern University, Dept. of Physics \&
  Astronomy, 2145 Sheridan Rd., Evanston, IL 60208}
}

\author{Wijnands, R.}{
  address={Astronomical Institute "Anton Pannekoek", University of Amsterdam, Kruislaan 403, 1098 SJ, The Netherlands}
}

\begin{abstract}
 The accreting millisecond pulsar SAX J1808.4-3658 may be a transition object 
between accreting X-ray binaries and millisecond radio pulsars.  We have 
constrained the thermal radiation from its surface through XMM-Newton X-ray
observations, providing strong evidence for neutrino cooling processes from 
the neutron star core.  We have also undertaken simultaneous X-ray and 
optical (Gemini) observations, shedding light on whether the strong heating 
of the companion star in quiescence may be due to X-ray irradiation, or to a 
radio pulsar turning on when accretion stops.
\end{abstract}


\maketitle


{\bf Introduction:}
The X-ray transient SAX J1808.4-3658 (hereafter 1808) has provided
many fundamental breakthroughs in the study of accreting neutron
stars (NSs); the first coherent millisecond X-ray pulsations 
discovered \citep{Wijnands98}, burst oscillations at the known spin 
frequency \citep{Chakrabarty03}, insight into the meaning of the 
frequency difference in kilohertz quasiperiodic oscillations 
\citep{Wijnands03c}.  

1808 has also provided two intriguing advances through study of its 
behavior in quiescence; its particularly low quiescent X-ray luminosity, 
and its relatively high optical luminosity in quiescence.  
We have made advances in understanding each issue, with implications 
for the nature of neutron star interiors and for the transition from 
X-ray binary to radio pulsar behavior.

\subsection{Quiescent X-ray Luminosities and Neutron Star Cooling}

Transiently accreting NSs in quiescence are usually seen to have soft,
blackbody-like X-ray spectra, often accompanied by
a harder X-ray component generally fit by a power-law of photon index
1-2 \citep[see Jonker, this volume; ][]{Campana98a}.  The harder component is of unknown origin;  
an effect of continued accretion, or a shock from a pulsar wind have 
been suggested \citep{Campana98a}.   
 The blackbody-like component is generally understood as the 
radiation of heat from the NS surface. This heat is produced by deep crustal 
heating during accretion, and is radiated by the crust 
on a timescale of $10^4$ years, producing a steady
quiescent thermal NS luminosity \citep{Brown98, Campana98a}. 
The deep crustal heating rate can be computed if the mass transfer rate is 
known (or estimated). 

However, the quiescent luminosity may be less than expected from 
``standard cooling'' if enhanced neutrino cooling processes are able 
to operate in the NS core.  The direct URCA process 
($n \rightarrow p+e+\bar{\nu}$, $p+e \rightarrow n + \nu$) is the simplest, 
and requires that protons constitute a significant component of the NS core, 
$>$10\% by mass.  Other rapid neutrino emission processes may involve 
hyperons, kaon-like condensates, and pion-like condensates.  All of 
these processes have sharp density thresholds, but are suppressed by 
proton superconductivity.  If proton superconductivity occurs at low densities 
and slowly turns off with increasing density, a range of cooling rates 
between ``standard'' cooling and the highest cooling rates are possible, for 
a range of NS masses \citep[see Yakovlev, this volume; ][]{Yakovlev04}.

Some transiently accreting NSs have been shown to have very low 
quiescent thermal X-ray luminosities \citep[e.g.][]{Jonker07}.
  This indicates enhanced neutrino emission from the core 
(or extremely long quiescent intervals, if the time-averaged mass 
transfer rate is unknown).  
Two transiently accreting NSs, 1808 and 1H 1905+000,
 provide the strongest constraints 
to date on neutrino cooling from NS cores, as a broader range of 
neutrino cooling rates is required from them than from young cooling pulsars 
\citep{Yakovlev04}. 

1808 has a well-determined distance of 3.4-3.6 kpc \citep{Galloway06}.  
Its mass transfer rate can be estimated as $10^{-11}$ \Msun/year, using 
the RXTE All-Sky Monitor count rates over the past 10 years (including 
4 outburst cycles).  
This is in remarkable agreement with the predictions of mass transfer by 
gravitational radiation of angular momentum in this system 
\citep{Bildsten01}, leading us to conclude that 1808's mass transfer rate 
(and thus crustal heating rate) is very well-known.  

A 2001 XMM-Newton observation of 1808 found an unexpectedly low quiescent 
X-ray luminosity for 1808, and an unexpectedly hard spectrum \citep{Campana02}. 
Deeper XMM observations in 2006 and 2007 confirmed this low quiescent 
luminosity ($L_X$(0.5-10 keV)=$5-8\times10^{31}$ ergs/s) and hard spectrum 
(which can be fit with a power-law of photon index 1.6 to 1.8).  
Fits with a power-law component plus a hydrogen-atmosphere model 
enables a constraint to be placed upon the temperature of any 
hydrogen-atmosphere model.  Simultaneous fits to all three XMM epochs 
allow a tight constraint of $kT<30$ eV, implying an unabsorbed bolometric 
$L_{NS}<5\times10^{30}$ ergs/s.  This luminosity is the lowest ever 
measured for the thermal component of any transient NS LMXB in quiescence. 
\footnote{Note that the {\it total} $L_X$ from 1H 1905+000 is lower still. } 


\begin{figure}
  \includegraphics[height=.35\textheight]{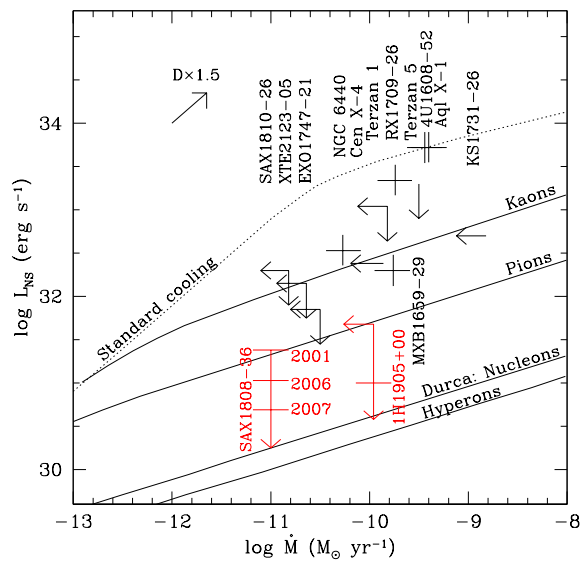}
  \caption{ \small{
Measurements of, or limits on, the quiescent thermal luminosity of 
various NS transients, compared to estimates of, or upper limits on, their 
time-averaged mass transfer rates.  Data from the compilation of 
\citet{Heinke07a}, plus \citet{Jonker07} and new work on 1808.  The predictions 
of standard NS cooling and models for enhanced cooling mechanisms are plotted 
following \citet{Yakovlev04}.  Multiple upper limits are plotted for 1808 and 
1H 1905+000. 
}
 }
\end{figure}

We have compiled the RXTE All-Sky Monitor lightcurves for 1808 and 10 other 
transient NS LMXBs, and used them to estimate their time-averaged 
mass transfer rates, or upper limits for those systems without a known 
outburst recurrence time.
  These mass transfer rates, along with the 0.1-10 keV 
thermal NS luminosities (or upper limits) are plotted in Figure 1, along 
with values for several other transient NS LMXBs from the literature 
\citep[see ][ for details]{Heinke07a}.  We give upper limits 
for 1808 and 1H 1905+000 \citep{Jonker07} 
from several observations, the most stringent 
coming from including 2007 observations of each.  

Predicted cooling curves for ``standard'' NS cooling, and for enhanced 
neutrino cooling processes involving protons (direct URCA), hyperons, 
kaons or pions, are plotted in Figure 1 in comparison with the data.  
The new measurements of 1808 and 1H 1905+000 are inconsistent with  
current models of cooling involving kaon or pion condensates, and suggest 
the presence of direct URCA losses through protons or hyperons.  

\subsection {1808's Unexplained Optical Luminosity}

The quiescent counterpart to 1808 was identified by \citet{Homer01} 
at $V$=21.5, brighter by five magnitudes than expected for a 
brown dwarf companion \citep{Bildsten01}.  The optical light was found 
by Homer et al. to be sinusoidally modulated at the orbital period, which 
is likely attributable to the varying aspect of the heated face of the 
secondary star.  However, the low quiescent X-ray luminosity of 1808 
may not be sufficient to produce the required irradiation of the secondary;
Homer et al. estimated that $L_{irr}>10^{33}$ ergs/s was required, while the 
X-ray observations have found $L_X<10^{32}$ ergs/s in quiescence.  
Since 1808 has shown some irregular variability during outbursts 
\citep{Wijnands01b}, we considered it important to observe 1808 
simultaneously in quiescence with X-ray and optical telescopes.  

We observed 1808 on March 10, 2007, with XMM-Newton and with the Gemini-South
 telescope, using the GMOS-S camera with the g' and i' filters.  
1808 was found to be in deep X-ray quiescence, with 
$L_X$(0.5-10 keV)$=8\times10^{31}$ ergs/s (extrapolation to 30 keV gives 
only $L_X=1.5\times10^{32}$ ergs/s).  
The very good seeing (0.65 to 0.98'') allowed us to resolve 1808's optical 
counterpart from a nearby (0.5'') star to the SE (g'=22.4), with which
 it is blended in \citet{Homer01} due to their poorer seeing.  Using 10 
uncrowded unsaturated nearby stars with the USNO B1.0 catalog, we find a 
position for 1808 of $\alpha$=18:08:27.63, $\delta$=-36:58:43.37 (J2000), with uncertainties of 0.2'' in each coordinate (accounting for the uncertainty in the transformation to the USNO B1.0 frame).  This is consistent with the 
VLA-derived position of \citet{Rupen02}, and the newly derived position of 
\citet{Hartman07}, while 1.7'' away from the position of \citet{Giles99}. 
We show our g' reference frame (made from 7 of the best seeing frames) in Figure 2. 

\begin{figure}
  \includegraphics[height=.29\textheight]{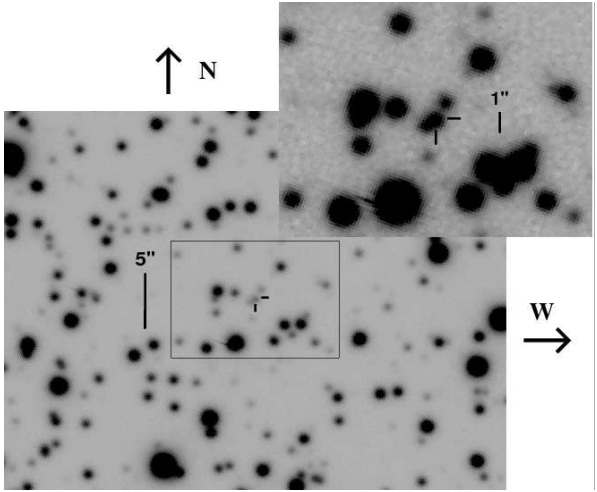}
  \caption{Finding chart for 1808, in g' from Gemini-S with 0.65'' seeing. 
A nearby (0.5'') star can be barely distinguished to the SE.}
\end{figure}

We find an average g' magnitude of 21.7, with a sinusoidal modulation 
(amplitude 0.27 magnitudes) that is consistent in phase with the 
visibility of the heated side of the companion (see Figure 3). 
 We find 1808 to have a larger (by a factor of five) orbital modulation 
than found by Homer et al.,  which may be partly attributed to our resolving 
1808 from nearby stars.  The total averaged optical luminosity
 in the 400-1000 nm range is estimated at $1.0\times10^{31}$ ergs/s,
 while the maximum to minimum luminosity variation is $6\times10^{30}$ ergs/s. 
  Our simultaneous observations prove that it is 
impossible to account for this variation by heating of the companion star 
with the observed X-ray luminosity of 1808; if a fraction $\sim0.011$ of the 
irradiating flux from the NS is intercepted by the companion 
\citep{Burderi03}, then an X-ray 
luminosity of $>5\times10^{32}$ ergs/s will be required to power this.  
We are currently modeling the observed data with a binary 
lightcurve synthesis code \citep{Orosz00} to determine what the contributions 
from the disk and irradiated companion star to the observed flux may be.

\begin{figure}
  \includegraphics[height=.35\textheight]{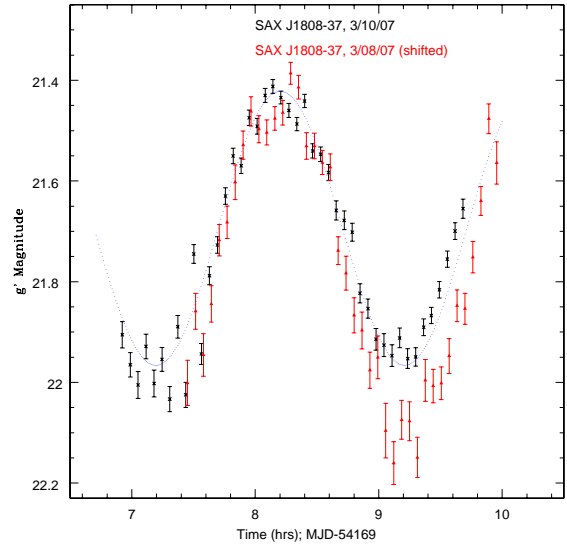}
  \caption{Orbital light curve for 1808, observed in g' band from Gemini-S. 
Observations from two nights are shown; black from March 10, 
grey from March 8 (in somewhat poorer seeing, 
shifted an integer number of orbits to match the March 10 
data).  A best-fit sinusoid has been plotted (dotted line) over the March 10 
data. }
\end{figure}

The disagreement between the quiescent X-ray luminosity and the sinusoidal 
optical modulation has been considered by \citet{Burderi03} and 
\citet{Campana04b}, who both concluded that the most likely source of the 
irradiating luminosity is a wind of relativistic particles associated 
with an active radio pulsar.  A similar disagreement between the quiescent 
X-ray luminosity and the amplitude of orbital modulation has now been 
observed for the similar accreting millisecond pulsar IGR J00291+5934 
\citep{D'Avanzo07}.  
Our simultaneous observation of 1808 in 
the X-ray and optical gives additional weight to these studies.  

However, it is difficult to understand how the radio pulsar, when active,  
would not evaporate the disk.  
The regularity of 1808's X-ray outbursts, and good agreement between the 
expected mass transfer rate from the companion and the time-averaged 
accretion rate onto the NS (see above), suggest that most material 
transferred from the companion reaches the NS.  
Another speculative possibility is a jet launched by 1808 in quiescence; 
a jet has been observed from 1808 in outburst \citep{Gaensler99}. 
If the NS's rotational axis is misaligned with the orbital axis, 
such a jet could impact the companion and may provide the 
necessary illumination; the feasibility of this possibility has 
not been thoroughly investigated.  
1808's unusually large optical luminosity and 
orbital modulations remain somewhat mysterious.

SAX J1808.4-3658 has been an invaluable laboratory for understanding the 
behavior of accreting NSs.  The unsolved problems associated with it 
hold the promise of unlocking additional facets of NS behavior.



\begin{theacknowledgments}
  We thank S. Sengupta for continuing conversations.  COH thanks A. Bonanos 
for assistance with the ISIS software.  
We thank M. Prakash, D. Page, K. Levenfish,  D. Yakovlev and D. Chakrabarty
 for discussions.  
 COH acknowledges the Lindheimer Postdoctoral Fellowship at 
Northwestern University.  COH and CJD acknowledge support from NASA 
XMM grant NNX06AH62G.  PGJ acknowledges support from the Netherlands 
Organization for Scientific Research.  CJD acknowledges support from NASA 
through the Chandra X-ray Center, grant number TM7-800.  
Based on observations obtained with XMM-Newton, an ESA science mission with instruments and contributions directly funded by ESA Member States and NASA. 
Based on observations obtained at the Gemini Observatory, which is operated by the Association of Universities for Research in Astronomy, Inc., under a cooperative agreement with the NSF on behalf of the Gemini partnership: The National Science Foundation (United States), the Particle Physics and Astronomy Research Council (United Kingdom), the National Research Council (Canada), CONICYT (Chile), the Australian Research Council (Australia), CNPq (Brazil), and CONICET (Argentina).

\end{theacknowledgments}


\bibliographystyle{aipproc}   


\bibliography{src_ref_list}

\end{document}